\documentclass[twocolumn,showpacs,preprintnumbers,amsmath,amssymb,aps,reprint]{revtex4-1}

\usepackage{graphicx}
\usepackage{epsfig}
\usepackage{dcolumn}
\usepackage{bm}
\usepackage{tabularx}

\newcommand{\bei}{\begin{itemize}}
\newcommand{\eei}{\end{itemize}}
\newcommand{\beq}{\begin{equation}}
\newcommand{\eeq}{\end{equation}}
\newcommand{\beqn}{\begin{eqnarray}}
\newcommand{\eeqn}{\end{eqnarray}}
\newcommand{\beqns}{\begin{eqnarray*}}
\newcommand{\eeqns}{\end{eqnarray*}}

\newcommand{\jprlBase}       {Phys.\ Rev.\ Lett.}
\newcommand{\jprBase}        {Phys.\ Rev.}
\newcommand{\jplBase}        {Phys.\ Lett.}

\newcommand{\npBase}         {Nucl.\ Phys.}

\newcommand{\aplet}     [1]  {{Appl.\ Phys.\ Lett.\ {\bf #1}}}

\newcommand{\npb}       [1]  {\npBase\ B~{\bf #1}}

\newcommand{\plb}       [1]  {\jplBase\ B~{\bf #1}}

\newcommand{\jprl}      [1]  {\jprlBase\ {\bf #1}}
\newcommand{\jprd}      [1]  {\jprBase\ D~{\bf #1}}

\begin{document}

\title{A search for non-virialized axionic dark matter}

\author{J. Hoskins}
\author{J. Hwang}\altaffiliation{Currently at Department of Physics, Sungkyunkwan University, Suwon, Gyeonggi-do 440-746, Republic of Korea.}
\author{C. Martin}
\author{P. Sikivie}
\author{N.S. Sullivan}
\author{D.B. Tanner}
\affiliation{University of Florida, Gainesville, Florida 32611}
\author{M. Hotz, L. J Rosenberg, G. Rybka, and A. Wagner}
\affiliation{University of Washington, Seattle, Washington 98195}
\author{S.J. Asztalos}\altaffiliation{Currently at XIA LLC, 31057 Genstar Rd., Hayward CA, 94544.}
\author{G. Carosi}
\author{C. Hagmann}
\author{D. Kinion}
\author{K. van Bibber}\altaffiliation{Currently at Naval Postgraduate School, Monterey CA, 93943.}
\affiliation{Lawrence Livermore National Laboratory, Livermore, California, 94550}
\author{R. Bradley}
\affiliation{National Radio Astronomy Observatory, Charlottesville, Virginia 22903}
\author{J. Clarke}
\affiliation{University of California and Lawrence Berkeley National Laboratory, Berkeley, California 94720}

\date{\today}

\begin{abstract}
Cold dark matter in the Milky Way halo may have structure defined by flows with low velocity dispersion. The Axion Dark Matter eXperiment high resolution channel is especially sensitive to axions in such low velocity dispersion flows. Results from a combined power spectra analysis of the high resolution channel axion search are presented along with a discussion of the assumptions underlying such an analysis. We exclude KSVZ axion dark matter densities of $\rho \gtrsim 0.2$ GeV/cm$^{3}$ and DFSZ densities of $\rho \gtrsim 1.4$ GeV/cm$^{3}$ over a mass range of $m_a = 3.3\mu$eV to 3.69$\mu$eV for models having velocity dispersions of $\Delta\beta\lesssim3\times10^{-6}$.
\end{abstract}

\maketitle

The Cold Dark Matter (CDM) component of the Milky Way halo may consist of low mass ($\mu$eV-meV) pseudoscalar particles called axions, which are a consequence of the Peccei-Quinn solution to the strong CP problem\cite{Peccei.77.1, Weinberg.78, Wilczek.78}. Axions of this mass range would have extremely weak couplings to standard-model particles as well as to both axionic and non-axionic dark matter, rendering them effectively collisionless\cite{Preskill.83, Abbot.83, Dine.83}. Models of the structure of halos consisting of such particles predict discrete flows with low velocity dispersion and high density at special locations called caustics\cite{Sikivie.92, Sikivie.03, Duffy.08}. In a Sikivie type detector\cite{Sikivie.83, Sikivie.85}, a low dispersion axion flow would appear as a narrow peak in the power spectrum. The Axion Dark Matter eXperiment (ADMX) high resolution (HR) data acquisition channel is sensitive to such signals, even more so than for signals expected from halo models lacking low dispersion flows\cite{Duffy.06}. Sensitivity to these signals is further improved by combining data from successive power spectra. This gain in the signal to noise ratio (SNR) relative to a single-scan analysis, such as in the previous ADMX HR analysis found in Ref. \cite{Duffy.06}, allows for sensitivity to lower densities, though care must be taken to account for the change in frequency of the axion signal due to the Earth's motion relative to the dark matter. Because any measurement of the axion's kinetic energy will be subject to Doppler shifts, both spectral broadening and signal modulation are expected. In this paper we discuss the constraints on a combined-spectra analysis imposed by Doppler shifts, and present the results of the first ADMX combined-spectra HR analysis.

ADMX converts axions to detectable microwave photons via the inverse Primakoff effect within a tunable, high quality factor ($Q\simeq50,000$) microwave cavity immersed in a strong magnetic field. An axion with mass $m_a$ and velocity $\beta$ would create a photon with energy $E \approx m_ac^2 + \frac{1}{2}m_ac^2\beta^2$. A velocity dispersion $\Delta\beta$ yields spectral broadening of $\frac{\Delta f}{f} = 2\frac{\Delta E}{E} \approx 2\beta \Delta\beta$. Axion conversion is expected to produce power in the cavity of \cite{Sikivie.83, Sikivie.85},

\begin{equation}
\label{eq:P}
P = g_{a\gamma\gamma}^2\frac{VB_0^2\rho_aC}{m_a}\min(Q,Q_a),
\end{equation}

\noindent where \emph{V} is the volume of the cavity, $B_0$ is the magnetic field strength, and $\rho_a$ is the axion density in the detector. The mode dependent form factor of the cavity is given by

\begin{equation}
\label{eq:C}
C = \frac{\left(\int_V\textbf{E}\cdot\textbf{B}_0d^3x\right)^2}{VB_0^2\int_V \left|\textbf{E}\right|^2d^3x},
\end{equation}

\begin{figure*}[t]
  \begin{center}
    \begin{tabular}{lr}
      \epsfig{file=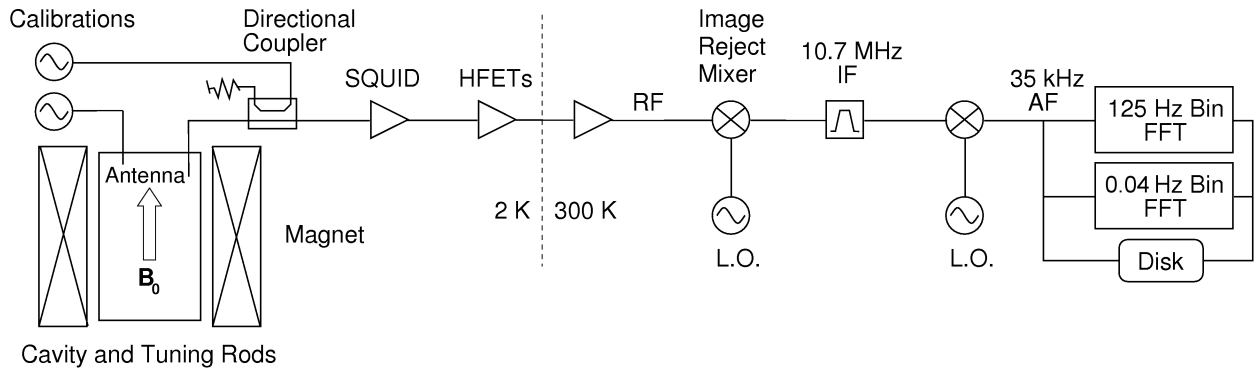,width=17.0cm}
    \end{tabular}
    \caption{\label{fig:Chain} Diagram of the ADMX cavity and receiver chain. The power read out from the cavity is sent through both cold and room temperature amplification stages, is mixed down from radio frequencies to audio frequencies via two local oscillators (L.O.), and is ultimately saved to disk.}
  \end{center}
\end{figure*}

\noindent where \textbf{E} is the electric field of the tuned mode and $\textbf{B}_0$ is the external magnetic field. The highest form factor obtainable (\emph{C} $\approx$ 0.69) corresponds to the TM$_{010}$ mode\cite{Peng} prompting its use over higher order modes. The quality factors of the cavity and the axions are $Q\lesssim10^5$ and $Q_a\gtrsim10^6$, respectively. Finally the axion to two photon coupling constant $g_{a\gamma\gamma}$ is given by

\begin{equation}
\label{eq:gagg}
g_{a\gamma\gamma}=\frac{\alpha g_\gamma}{\pi f_a},
\end{equation}

\noindent where $\alpha$ is the fine structure constant, $f_a$ is the Peccei-Quinn symmetry breaking scale, and $g_\gamma$ is a model dependent constant. In the Kim-Shifman-Vainshtein-Zakharov (KSVZ) model $g_\gamma=-0.97$\cite{Kim.79, Shifman.80}, and in the Dine-Fischler-Srednicki-Zhitnitskii (DFSZ) model $g_\gamma=0.36$\cite{Dine.81, Zhit.80}.

The ADMX detector consists of a 1-m tall, 0.5-m diameter, copper plated, stainless steel, right cylindrical cavity kept at 1.8 K and placed in a 7.6 T magnetic field. The TM$_{010}$ mode of the cavity is swept through a range of frequencies, corresponding to a range of possible axion masses, by moving a metal rod, oriented parallel to the cylinder axis and extending along the full length of the cavity, from the wall to the center of the cavity space\footnote{There are two such rods for redundancy, and for tuning past mode degeneracies in the cavity.}. For the experimental parameters of ADMX, Eq.~\ref{eq:P} predicts an expected signal power of order 10$^{-22}$ W. The expected \emph{SNR} is given by the Dicke radiometer equation \cite{Dicke}:

\begin{equation}
\label{eq:SNR}
SNR = \frac{P}{P_N}\sqrt{bt} = \frac{P}{k_BT_N}\sqrt{\frac{t}{b}},
\end{equation}

\noindent where \emph{P} is the expected signal power, $P_N$ is the noise power, \emph{b} is the signal bandwidth, \emph{t} is the integration time, $k_B$ is Boltzmann's constant, and $T_N$ is the total noise temperature which is equal to the physical temperature plus the noise temperature of the electronics.

ADMX has two data acquisition channels which use the receiver and amplifier chain shown in Fig.~\ref{fig:Chain}. Power from the cavity is critically coupled to the receiver which sends the signal through a directional coupler followed by two cold ($\simeq$ 1.8 K) amplification stages and then one room temperature amplification stage. The cold amplification stages are a DC Superconducting QUantum Interference Device (SQUID) amplifier\cite{Mück} followed by two balanced GaAs heterostructure field-effect transistor (HFET) amplifiers. There is $\sim$10 dB gain from the SQUID and $\sim$ 34 dB combined gain from the HFETs\cite{Daw.97, Bradley.03, ADMX.11}. At room temperature, the signal is amplified an additional 35 dB before being down-converted to a center frequency of 10.7 MHz and sent through an eight-pole crystal band pass filter. The signal is down-converted a second time to a center frequency of 35 kHz whereupon the medium resolution (MR) and HR channels diverge.

The MR channel takes 10,000 scans over 80 seconds, computes a fast Fourier transform (FFT) and power spectrum of each scan and averages the 10,000 spectra into one 400-point power spectrum with a resolution of 125 Hz. The HR channel takes data for 23.8 seconds with a sampling rate of 80 kHz resulting in a maximum resolution of 42 mHz. Three such spectra are taken during the integration time of the MR channel. The HR data are saved in two formats. First the raw time series data are written to disk with no signal processing or averaging having been applied. Analysis of these data will be the subject of a future publication. Second, the three scans taken during the MR integration are Fourier transformed, averaged, and saved as a single 42-mHz bin-width-power spectrum. After acquiring data at a given resonant frequency, $f_0$, the tuning rods are moved, shifting $f_0 \sim 2$ kHz, and the data acquisition process is repeated. Because the tuning rod steps are much smaller than the 30-kHz bandwidth of the crystal filter and the resonant width of the cavity TM$_{010}$ mode, the frequency coverage for successive scans overlap significantly. For both channels, the total exposure time for a given frequency bin is $\sim$25 minutes.

\begin{figure}
  \begin{center}
    \begin{tabular}{lr}
      \epsfig{file=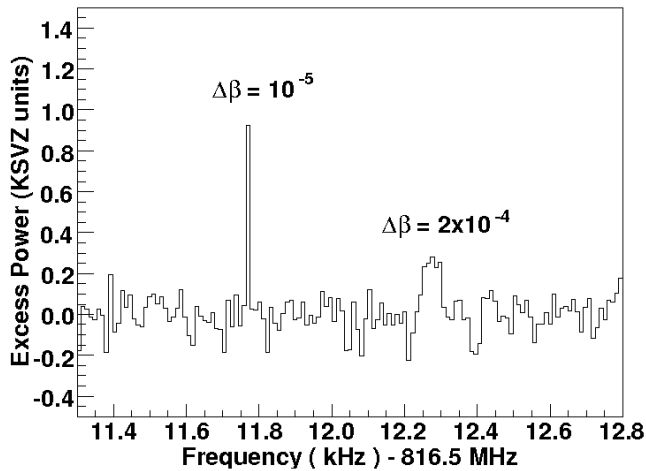,width=8.6cm}
    \end{tabular}
    \caption{\label{fig:Sim} Monte Carlo simulations of non-virialized axion signals imposed on real HR data. The broader peak (velocity dispersion of $2\times10^{-4}$) is shown as an example of the narrowest signal that the MR channel can resolve. An excess power of 0 corresponds to power in that bin equal to the rms noise power at that frequency.}
  \end{center}
\end{figure}

The MR channel has recently produced limits for both virialized and non-virialized axions with $\Delta\beta \sim 10^{-3}$ and $\lesssim 2\times10^{-4}$, respectively\cite{ADMX.09}. While the MR and the HR channels both look for non-virialized axions, the difference comes in their sensitivity to models with differing velocity dispersions. The MR channel is equally sensitive to all signals with dispersions of $\lesssim10^{-4}$, as all such signals would occupy a single 125 Hz bin. The HR channel potentially has a much smaller bin width, and therefore continues to gain in the SNR for signals with low velocity dispersion. Figure~\ref{fig:Sim} shows the expected signals in the HR channel for two simulated axion flows of equal density but differing velocity dispersion. While both signals would be be seen in the HR channel, the signal having lower velocity dispersion stands dramatically above the noise. Note that at its highest resolution, the HR channel would have non-Gaussian noise. For a detailed discussion of the possible origins and the noise characteristics of low dispersion axion signals see Duffy \textit{et al}.\cite{Duffy.06}.

A combined-spectra analysis relies on the signal adding coherently while the noise adds randomly. Thus it is most effective if the signal remains in the same bin from spectrum to spectrum. Because the HR channel is affected by the Doppler shift of an axion signal due to the time varying velocity of the detector, this constraint imposes a lower limit on the bin width of this analysis. In a worst-case scenario, the combined daily signal modulation from both the orbital and rotational motions of the Earth is at most a few hertz\cite{Turner.90, ling.04}. A bin width larger than twice the modulation amplitude is required to minimize the probability of the signal drifting into a neighboring bin between scans. Therefore, each neighboring set of 256 bins in the 42 mHz spectrum was co-added, yielding a spectrum with a bin width of 10.8 Hz. The resulting spectrum is sufficiently averaged so as to possess Gaussian noise and to be mostly insensitive to the signal modulation caused by both orbital and rotational terrestrial motion on the time scale of a few days.

\begin{figure}
  \begin{center}
    \begin{tabular}{lr}
      \epsfig{file=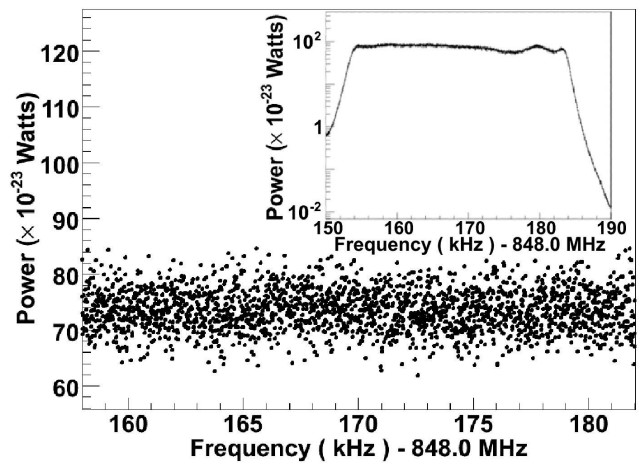,width=8.5cm}
    \end{tabular}
    \caption{\label{fig:Raw} An example of a modulation-insensitive power spectrum after the crystal filter shape and amplifier chain effects have been divided out. The insert shows the raw power spectrum with a passband shape imparted by the crystal filter.}
  \end{center}
\end{figure}

The modulation-insensitive spectra, shown in figure~\ref{fig:Raw}, were corrected for systematic effects, the most noticeable being the shape given to the power spectrum by the crystal filter. A reference spectrum representative of the crystal filter shape was created by averaging several days worth of data. This reference spectrum was divided out before each spectrum was cropped to the 30-kHz bandwidth of the crystal filter. Any lingering broad spectral structure was due to frequency dependent interactions within the amplifier chain. A sixth-order polynomial was fit to and divided out of each spectrum to remove this structure.

Frequencies with excessive power (i.e., bins containing power which is comparable to the expected power deposition from KSVZ axions) and/or excessive noise (i.e., SNR below 5) were rescanned within a day or two. These rescans were averaged with the previous data set to reduce noise caused by statistical fluctuations. For frequencies having too much power to exclude a KSVZ axion signal at 90\% confidence, rescans totaling an additional 10 to 20 minutes of exposure time were taken to determine if the signal warranted further consideration as an axion candidate. All rescans extended above and below the frequency in question by several kHz. Because an actual axion signal must be persistent, the disappearance of a signal would indicate that its origin is either from statistical fluctuations or from an external transient source, rather than from an axion flow. In the frequency range of 812.0 MHz to 892.8 MHz (3.3 $\mu$eV - 3.69 $\mu$eV) we rescanned several frequencies with excess power, though no signals persisted through the rescan process.


\begin{figure}
  \begin{center}
    \begin{tabular}{lr}
      \epsfig{file=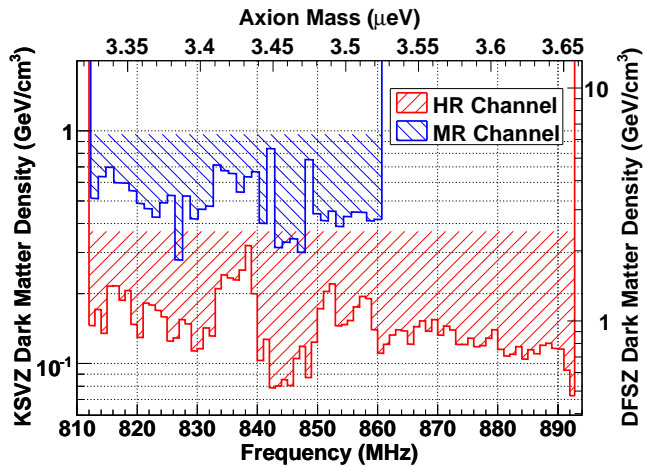,width=8.5cm}
    \end{tabular}
    \caption{\label{fig:Lim} Density limits for the HR channel ($\Delta\beta \lesssim 3\times10^{-6}$) at 90\% confidence from 812 MHz to 892.8 MHz. The scale for limits on KSVZ and DFSZ axions are shown on the left and right axes respectively. Density limits for the currently published MR channel data ($\Delta\beta \lesssim 2\times10^{-4}$)\cite{ADMX.09} are shown for comparison.}
  \end{center}
\end{figure}

Because no signals remained after performing rescans, the combined \emph{SNR} and accumulated power were used to set limits on the local density of non-virialized axionic dark matter. Both KSVZ and DFSZ models are considered. In the 800 MHz range, a peak in the power spectrum from an axion flow with velocity relative to the detector of $\beta \sim 10^{-3}$ and dispersion of $\Delta\beta \lesssim 3\times10^{-6}$ would have a width of $\lesssim$ 5 Hz. The limits presented here are for axion flows having this dispersion or lower. At 90\% confidence, we limit the local density of non-virialized axions to exclude $\rho \gtrsim 0.2$ GeV/cm$^{3}$ for KSVZ models and $\rho \gtrsim 1.4$ GeV/cm$^{3}$ for the DFSZ model. Taking 0.52 GeV/cm$^{3}$ as an estimate of the average local density\cite{Gates, Salucci.10}, non-virialized axions are excluded for KSVZ models over a mass range of $m_a = 3.3 \mu$eV to 3.69 $\mu$eV. Consider also the caustic ring halo model, presented in detail in Ref. \cite{Duffy.08}, which predicts that most of the dark matter in the local neighborhood is in discrete flows. At the Earth's location in the Milky Way, one such flow is predicted to have a density as large as $\rho \approx 0.84$ GeV/cm$^{3}$ which is well over our stated exclusion limit for KSVZ model axions. The HR density limits for non-viralized KSVZ and DFSZ axions are shown in Fig.~\ref{fig:Lim} with the MR limit (calculated from \cite{ADMX.09}) overlayed for comparison.

The existing time-series data will permit a future analysis of this same data set that will be capable of reaching finer resolutions while still accounting for terrestrial signal modulation. The analysis will return to a single initial scan analysis, where candidate signals found in one scan will be compared against any rescans having frequency coverage corresponding to where the signal could have moved to as a result of Doppler shifts. The time between scans will determine the maximum expected signal modulation for these candidates. To warrant further consideration, a candidate signal must be present in rescans within a window set by the maximum Doppler modulation about the original candidate frequency. In addition, taking the FFT of smaller intervals of the time series data enables a range of dispersions to be tested. Further, it can be seen in Eq.~\ref{eq:SNR} that the SNR scales inversely as the square root of the signal bandwith. Thus a finer resolution analysis would yield a higher sensitivity to $g_{a\gamma\gamma}$. We expect approximately an order of magnitude improvement in sensitivity for the HR channel using such an analysis at a resolution of 100 mHz.

This work has been supported by the U.S. Department of Energy through grant numbers DE-FG02-97ER41029, DE-FG02-96ER40956, DE-AC52-07NA27344, and DE-AC03-76SF00098, as well as through the LDRD program at Lawrence Livermore National Laboratory.



%

\end{document}